\documentclass[10pt,twocolumn,twoside]{IEEEtran}
\usepackage{color}
\usepackage{cite}
\usepackage{epsfig}
\usepackage{amssymb}
\usepackage{amsmath}
\usepackage{amsthm}
\usepackage{booktabs}
\usepackage{multirow}
\usepackage{hyperref}

\newcommand{\as}{a}
\newcommand{\bs}{b}
\newcommand{\cs}{c}

\newcommand{\Pmd}{\mathbb{P}_{\mathrm{md}}}
\newcommand{\Pfa}{\mathbb{P}_{\mathrm{fa}}}
\newcommand{\Pv}{\mathbb{P}_v}
\renewcommand{\Pi}{\mathbb{P}_\infty}
\newcommand{\Po}{\mathbb{P}_1}
\newcommand{\E}{\mathrm{E}}

\newcommand{\ma}{m_\alpha}
\newcommand{\Tf}{T_{\mathrm{F}}}

\newcommand{\Fo}{F_{z,0}}
\newcommand{\Ho}{\mathcal{H}_0}
\newcommand{\Hu}{\mathcal{H}_1}
\newcommand{\LLR}{\mathrm{LLR}}
\newcommand{\N}{\mathcal{N}}

\newcommand{\Xs}{X_n}
\newcommand{\Ys}{Y_n}

\newcommand{\sigssu}{\sigma_1^2}
\newcommand{\sigsso}{\sigma_0^2}
\newcommand{\musu}{\mu_1}
\newcommand{\muso}{\mu_0}

\newcommand{\sst}{\sigssu}
\newcommand{\mt}{\musu}

\newcommand{\tz}{\tilde{z}}

\newcommand{\sigz}{\sigma_z}

\newcommand{\zt}{\tilde{z}}

\newcommand{\bedg}{\beta_{\mathrm{EDG}}}
\newcommand{\bclt}{\beta_{\mathrm{CLT}}}
\newcommand{\aedg}{\alpha_{\mathrm{EDG}}}
\newcommand{\aevt}{\alpha_{\mathrm{EVT}}}

\newcommand{\p}[1]{\left(#1\right)}
\newcommand{\corch}[1]{\left[#1\right]}
\newcommand{\llav}[1]{\left\{#1\right\}}
\newcommand{\eq}[1]{\begin{equation}{#1}\end{equation}}
\newcommand{\hyp}[6]{\eq{
#1\sim \begin{cases}
\Ho :\quad \N\p{#2,#3} & \mbox{if } 1\leq n < v \mbox{ or } n \geq v+m\\
\Hu :\quad \N\p{#4,#5} & \mbox{if } v\leq n < v+m
\end{cases}#6}}

\newtheorem{col}{Corollary}
\newtheorem{theorem}{Theorem}
\newtheorem{lemma}{Lemma}
\newtheorem{prop}{Proposition}

\begin{document}
%
\title{Closed-Form Approximations for the Performance Upper Bound of Inhomogeneous Quadratic Tests}

\author{Daniel~Egea-Roca,~\IEEEmembership{Student Member,~IEEE,}
		Gonzalo~Seco-Granados,~\IEEEmembership{Senior Member,~IEEE,}
		and~Jos\'e~A.~L\'opez-Salcedo,~\IEEEmembership{Senior Member,~IEEE}

\thanks{This work was partly supported by the Spanish Government under grant
TEC2014-53656-R.}}

\maketitle
\begin{abstract}
This paper focuses on inhomogeneous quadratic tests, which involve the sum of a dependent non-central chi-square with a Gaussian random variable. Unfortunately, no closed-form expression is available for the statistical distribution of the resulting random variable, thus hindering the analytical characterization of these tests in terms of probability of detection and probability of false alarm. In order to circumvent this limitation, two closed-form approximations are proposed in this work based on results from Edgeworth series expansions and Extreme Value Theory (EVT). The use of these approximations is shown through a specific case of study in the context of integrity transient detection for Global Navigation Satellite Systems (GNSS). Numerical results are provided to assess the goodness of the proposed approximations, and to highlight their interest in real life applications.
\end{abstract}


\section{Introduction}
\label{sec:Intro}

Detecting the presence of an event is a recurrent problem in several fields such as medicine, finance, speech processing or radar, just to mention a few. Typically, decisions are taken based on the value provided by some function of the observed data, which is compared to a predefined threshold for accepting or rejecting the hypothesis under analysis (e.g. the event is present or not). This function of the observed data is often referred to as the {\em test statistic} for the problem at hand, and it can be obtained applying different optimization criteria. For instance, the well-known Neyman-Pearson (NP) criterion, which aims at maximizing the probability of detection subject to some probability of false alarm; the Generalized Likelihood Ratio Test (GLRT), which replaces unknown parameters by their maximum likelihood estimates (MLE); or the Bayesian criterion, which aims at minimizing a weighted function of different error probabilities, known as the Bayesian risk \cite{levy}.

When it comes to assess the detection performance, the first step is to determine the statistical distribution of the test statistic, and in particular, its cumulative density function (cdf). This allows the designer to assess the receiver operating characteristics (ROC) based on the probability of (missed) detection and probability of false alarm, thus having a full picture of the overall detection performance.

In many applications dealing with Gaussian distributed data, the resulting test statistic is based on homogeneous quadratic forms. That is to say, quadratic forms that are composed of a linear combination of quadratic terms or a mixture of both quadratic and crossed terms. Unfortunately, a closed-form expression is difficult to be obtained for the general class of homogeneous quadratic forms, since the presence of linearly combined correlated terms often makes the problem mathematically intractable \cite{johnson70,rice80}. This is the case of tests statistics based on the weighted sum of chi-squared random variables \cite{bodenham15}, which have received significant attention in the past decades in the context of spacecraft engineering \cite{feiveson68}, goodness-of-fit tests \cite{zhang05}, multiuser interference in broadcast channels \cite{jang07}, or cooperative spectrum sensing \cite{ma08}. Because of the difficulty in characterizing the linear combination of quadratic forms, the resulting distribution is numerically computed through approximate methods such as the saddle-point approximation \cite{wood93} or by matching a few of the first cumulants to some other known closed-form distributions \cite{buckley88}.

The problem is further aggravated when dealing with inhomogeneous quadratic tests, where apart from a combination of quadratic forms, linear terms are also present thus making even more difficult to characterize the overall statistical distribution. Inhomogeneous quadratic forms appear in applications such as neuron receptive fields modeling \cite{berkes07}, financial problems dealing with portfolio losses of CDO pricing \cite{schlosser}, or signal-quality monitoring in Global Navigation Satellite Systems (GNSS) \cite{ION}, just to mention a few. Nevertheless, the use of inhomogeneous quadratic forms has typically remained in the realm of estimation or optimization theory, where some parameters need to be estimated or where the forms are part of some optimization criterion \cite{stepniak11}. To the best of the authors' knowledge, little attention has been paid so far to inhomogeneous quadratic tests in detection problems, where the statistical characterization of these forms becomes of paramount importance. 

Motivated by this observation, the problem addressed in this paper is that of finding a closed-form approximation to the cdf of inhomogeneous quadratic forms, in order to easily compute the error probabilities in detection problems. The main goal is to obtain a compact and analytical formulation that can easily be parameterized in time-varying working conditions. This is an important requirement in practical applications such as integrity monitoring in GNSS, where the propagation conditions may vary quite rapidly, and the user receiver needs to promptly react for detecting any potential threat. The approach considered herein is based on the use of the Edgeworth series expansion, which provides an analytical expression for the probability density function (pdf) under analysis based on its constituent moments \cite[p.~169]{Kendall}. In contrast to other methods, such as the saddlepoint approximation, the advantage of the Edgeworth expansion is that is can easily be integrated to give an analytical expression for the cdf, and does not require the cumulant generating function to be known in closed-form \cite{nielsen79}.

The application of these results will be illustrated in a specific case study dealing with transient change detection (TCD) in the context of GNSS signal-level integrity. The goal is to detect abrupt changes of finite duration in the time evolution of quality monitoring metrics, and in particular, those modeled by inhomogeneous quadratic forms. We will see how the Edgeworth series expansion is accurate enough for approximating the probability of miss, but not that much when dealing with the probability of false alarm. In that case, we will propose an alternative approximation based on results from Extreme Value Theory (EVT), which again makes use of the pdf and cdf of the inhomogeneous quadratic form under study. The use of Edgeworth and EVT provides a dual approach for assessing the detection performance of inhomogeneous quadratic tests, allowing the reader to easily extend these results to a wide range of different applications other than the one considered herein.

The paper is organized as follows. Some preliminaries are first discussed in Section \ref{sec:Prel}, where the signal model for inhomogeneous quadratic tests is provided as well as some stepping-stone results. Next, the case study on TCD is introduced in Section \ref{sec:sigModel}. Closed-form approximations for the probability of miss and probability of false alarm are provided in Section \ref{sec:Edge} and Section \ref{sec:EVT}, making use of Edgeworth series and EVT, respectively. The use of these closed-form approximations on the specific application of GNSS signal-level integrity is discussed in Section \ref{sec:GNSS}, and finally, conclusions are drawn in Section \ref{sec:conclusions}.

\section{Preliminaries\label{sec:Prel}}
Let $\left\{Y_n\right\}_{n\geq 1}$ be a sequence of iid random variables whose inner structure is given by the following inhomogeneous quadratic form, 
\begin{equation}
Y_n\doteq aX_n^2+bX_n+c\label{defY}
\end{equation}
for some constants $\left\{a,b,c\right\}$ and $X_n$ some random variable. Because of the presence of a quadratic and a linear term in (\ref{defY}), the distribution of $Y_n$ is not straightforward. This problem is further aggravated when we actually intend to find the distribution for the sum of $\left\{Y_n\right\}_{n\geq 1}$, denoted herein by the random variable $Z$ as follows,
\begin{equation}
Z\doteq\sum_{n=1}^m Y_n\label{defZ}
\end{equation}
The distribution of $Z$, denoted by $f_Z$, involves $m$ times the convolution of the pdf of $Y_n$. This poses insurmountable obstacles for the derivation of a closed-form expression for $f_Z$, and therefore for the corresponding cdf, $F_Z$. In order to circumvent this limitation, we propose the use of some approximations that provide a tight match to the original distribution, while providing a mathematically tractable closed-form expression. We will briefly recall here the Central Limit Theorem, which becomes a simple reference benchmark for the approximations to be proposed later on, as well as some indications on the statistical moments of $Z$ to be used as well. 

\subsection{Central limit theorem (CLT) for the distribution of $Z$}
\begin{theorem}[{CLT}]
\label{th:CLT}
Let $Z$ be the sum of $m$ independent random variables $Y_1,Y_2,\ldots,Y_m$, with mean $\mu_Z=\mu_1+\mu_2+\ldots+\mu_m$ and variance $\sigma_Z^2=\sigma_1^2+\sigma_2^2+\ldots+\sigma_m^2$. Then,
\begin{eqnarray}
f_Z(\tilde{z})&\xrightarrow[m\rightarrow\infty]{}&\phi(\tilde{z})\doteq\frac{1}{\sqrt{2\pi}}e^{-\tilde{z}^2/2}\label{eq:CLT_PDF}\\
F_Z(\tilde{z})&\xrightarrow[m\rightarrow\infty]{}&\Phi(\tilde{z})\doteq\frac{1}{\sqrt{2\pi}}\int_{-\infty}^{\tilde{z}}e^{-\lambda^2/2}d\lambda\label{eq:CLT_CDF}
\end{eqnarray}
where $\tilde{z}\doteq (z-\mu_Z)/\sigma_Z$, $\phi(\tilde{z})$ the standard Gaussian pdf and $\Phi(\tilde{z})$ the standard Gaussian cdf \cite{Papoulis}.
\end{theorem}

\subsection{Moments of $Z$\label{secMomentsZ}}
\begin{lemma}
\label{prop:Z}
Let $Z$ be the sum of $m$ independent random variables $Y_1,Y_2,\ldots Y_m$. The $k$-th order moment of $Z$ denoted by $\xi_{Z}^k\doteq\mathrm{E}\left[Z^k\right]$ can be computed using the multinomial theorem \cite{bolton68} as follows,
\begin{equation}
\xi_{Z}^k=\sum_{l_1+l_2+\ldots+l_m=k}
\frac{k!}{l_1!l_2!\cdots l_m!}\xi_{Y_1}^{l_1}\xi_{Y_2}^{l_2}\cdots \xi_{Y_m}^{l_m}\label{multTh}
\end{equation}
for all sequences $\left\{l_n\right\}_{n=1}^m\in\mathbb{Z}^*$ such that their sum is equal to $k$, and where $\xi_{Y}^l$ stands for the $l$-th order moment of the inhomogeneous quadratic form in (\ref{defY}). 
\end{lemma}

The computation of moments up to order four will be used later in this work, so it is convenient to provide here the particular case of (\ref{multTh}) for $k=\left\{1,2,3,4\right\}$. After some cumbersome but straightforward manipulations, we have:
\begin{eqnarray}
\xi_Z&{}={}&m\xi_Y\\
\xi_{Z}^2&{}={}& m\left[\xi_{Y}^2+(m-1)(\xi_{Y})^2\right]\label{eq:mzs}\\
\xi_{Z}^3 &{}={}& m\Big[\xi_Y^3 + (m-1)\left[3\xi_Y^2\xi_Y +(m-2)(\xi_Y)^3\right]\Big]\\
\xi_{Z}^4 &{}={}& m\Big[\xi_Y^4 + (m-1)\big[4\xi_Y^3\xi_Y+3\xi_Y^2\nonumber\\
&&{+}\:(m-2)\big[6\xi_Y^2(\xi_Y)^2+(m-3)(\xi_Y)^4\big]\big]\Big].
\end{eqnarray}

\subsection{Moments of $Y_n$\label{secMomentsY}}
The moments of the inhomogeneous quadratic form in (\ref{defY}) could be obtained applying the multinominal theorem in (\ref{multTh}), as well, since the problem at hand is also a sum of random variables. However, having the sum of only three terms we can simplify (\ref{multTh}) into a more compact expression as follows:
\begin{lemma}
\label{prop:Y}
Let $\Ys \doteq \as\Xs^2 + \bs\Xs + \cs$ for some constants $\left\{a,b,c\right\}$ and some random variable $X_n$. Then the moments of $\Ys$ are given by
\eq{\label{eq:mys}
\xi_Y^k \doteq \E[\Ys^k] = \sum_{i=0}^k A(i),}
with
\eq{\label{eq:Ai}
A(i) \doteq \sum_{j=0}^i \binom{k}{i}\binom{i}{j} \as^{k-i} \bs^{i-j} \cs^j \xi_{X,2k-i-j},
}
where $\xi_X^k \doteq \E[\Xs^k]$ is the $k$-th order moment of $\Xs$ and $\binom{l}{i} \doteq k!/(i!(k-i)!)$ is the binomial coefficient.
\end{lemma}
\begin{IEEEproof}
In order to find a general expression for the moments of $\Ys$ we rewrite $Y_n$ as
\eq{
\Ys^l = \p{\as\Xs^2 + \bs\Xs + \cs}^k = (u+v)^k,
}
with $u \doteq \as\Xs^2$ and $v \doteq \bs\Xs + \cs$. Applying the binomial expansion we have,
\eq{
\Ys^k = \sum_{i=0}^k \binom{k}{i} u^{k-i}v^i = \sum_{i=0}^k\binom{k}{i}\as^{k-i}\Xs^{2(k-i)}\p{\bs\Xs+\cs}^i.\label{tmp0}
}
Applying again the binomial expansion to the right-hand side of (\ref{tmp0}), we get
\eq{
\begin{aligned}
\Ys^k &= \sum_{i=0}^k \binom{k}{i}\as^{k-i}\Xs^{2(k-i)} \p{\sum_{j=0}^i \binom{i}{j}(\bs\Xs)^{i-j}\cs^j} \\
&= \sum_{i=0}^k \sum_{j=0}^i \binom{k}{i}\binom{i}{j} \as^{k-i} \bs^{i-j} \cs^j \Xs^{2k-i-j}.
\end{aligned}}
Thereby, by the definition of the moments of $\Ys$, $\xi_Y^k\doteq \E[\Ys^k]$, and $A(i)$ in \eqref{eq:Ai}, then \eqref{eq:mys} follows.
\end{IEEEproof}

\section{Case study on transient change detection\label{sec:sigModel}}
\subsection{Signal model}
Before addressing the proposed closed-form approximations, let us first introduce the signal model for the case study of TCD under consideration. This will unveil the need for an alternative to the Edgeworth series expansion, as far as probability of false alarm is concerned. To do so, let $\left\{X_n\right\}_{n\geq 1}$ be a sequence of random samples that are observed sequentially. We consider a family $\{\Pv | v\in [1,2,\dots,\infty]\}$ of probability measures, such that under $\Pv$, the observations $X_1,\dots, X_{v-1}$ and $X_{v+m},\dots, X_\infty$ are iid with a fixed marginal pdf $f_{X,0}$, with $v$ the deterministic but unknown change time when a change in distribution appears. On the other hand, $X_v,\dots, X_{v+m-1}$ are iid with another marginal pdf $f_{X,1}\neq f_{X,0}$. In our case, we focus on the simultaneous change of mean and variance on a Gaussian distribution,
\hyp{\label{eq:model}\Xs}{\muso}{\sigsso}{\musu}{\sigssu}{,}
where $\{\mu_0,\sigma_0^2\}$ are the mean and variance of $X_n$ under nominal conditions (i.e. hypothesis $\Ho$) and $\{\musu,\sigssu\}$ the mean and variance during the change (i.e. hypothesis $\Hu$). 

\subsection{Test statistic}
The detection of a transient change is completely defined by the \emph{stopping time} $T$ at which the change is declared, which can be computed following different rules and criteria \cite[Ch. 6]{zoubir}. In this work we will focus on the finite moving average (FMA) stopping time introduced in \cite{guepie12} for the specific case of Gaussian mean-changes, and recently extended to the case of Gaussian mean- and variance-changes in \cite{SSP}. The FMA test statistic results in the following stopping time
\eq{\label{eq:Tfma}
\Tf \doteq \inf\llav{n\geq m: S_n\geq h},
}
with $h$ the detection threshold and 
\eq{
S_n \doteq \sum_{k=n-m+1}^n \LLR_k\label{defSn},
}
where $\LLR_k\doteq f_{X,1}(X_k)/f_{X,0}(X_k)$ is the likelihood ratio (LLR) of the sample $X_k$. Interestingly, the LLR turns out to be an inhomogeneous quadratic form when evaluated for the signal model in (\ref{eq:model}). That is,
\begin{equation}
\mathrm{LLR}_k=aX_k^2+bX_k+c\label{defLLR}
\end{equation}
where the constants $\left\{a,b,c\right\}$ are given by:
\begin{eqnarray}
a &=& \frac{\sst - \sigsso}{2\sigsso\sst},\label{eq:param}\\
b &=& \frac{\sigsso\mt - \sst\muso}{\sigsso\sst},\\
c &=& \ln\p{\frac{\sigsso}{\sst}} + \frac{\sst\muso^2 - \sigsso\mt^2}{2\sigsso\sst}.
\end{eqnarray}
The detection metric in (\ref{defSn}) is actually the accumulation of $m$ inhomogeneous quadratic forms, and therefore it can be modeled by the random variable $Z$ in (\ref{defZ}). While the exact distribution of (\ref{defSn}) is unknown, the statistical moments can be derived using the results from Lemma \ref{prop:Z} and \ref{prop:Y}.

\subsection{Detection performance}
The detection performance is measured in terms of the worst-case probability of missed detection and false alarm, which are defined respectively, as 
\begin{eqnarray}
\Pmd(\Tf) &\doteq& \sup_{v\geq 1} \Pv(\Tf > v+m-1 |\Tf\geq v)\label{defPmd},\\
\Pfa(\Tf) &\doteq& \sup_{l\geq 1} \Pi(l\leq \Tf < l+\ma)\label{defPfa},
\end{eqnarray}
where $m$ is the length of the transient to be detected and $\ma$ the time window where we want a given value of $\Pfa$ to be guaranteed. The exact computation of (\ref{defPmd}) and (\ref{defPfa}) leads to an intractable formulation. To circumvent this limitation, upper bounds are typically adopted instead such that \cite{SSP},
\begin{eqnarray}
\Pmd(\Tf) & \leq & \beta_m(h), \\
\Pfa(\Tf) & \leq &\alpha_{\ma}(h).\label{upPfa1}
\end{eqnarray}
with
\begin{eqnarray}
\beta_m(h) 		&\doteq& \Po(S_n < h),\label{eq:beta}\\
\alpha_{\ma}(h) 	&\doteq& 1 - \corch{\Pi(S_n < h)}^{\ma}.\label{eq:alpha}
\end{eqnarray}
Due to the one-to-one relationship between (\ref{defSn}) and (\ref{defZ}), we can reformulate the upper bounds in (\ref{eq:beta})-(\ref{eq:alpha}) as
\begin{eqnarray}
\beta_m(h) &=& F_{Z,1}(h)\label{eq:beta2}\\
\alpha_{\ma}(h) &=& 1-\left[F_{Z,0}(h)\right]^{m_\alpha}\label{eq:alpha2}
\end{eqnarray}
with $F_{Z,0}$ and $F_{Z,1}$ the cdf of $Z$ in the absence and in the presence of a transient change, respectively.

These cdf have no closed-form expression either, but tight approximations can adopted instead. For instance, using the Edgeworth series expansion to be presented next in Section \ref{sec:Edge}. This approach works well for $\beta_m(h)$ in (\ref{eq:beta2}), since it directly depends on the cumulative density function of $Z$, for which the Edgeworth expansion can readily be derived using the moments of $Z$ introduced in Section \ref{secMomentsZ}. However, some difficulties are found for $\alpha_{\ma}(h)$ in (\ref{eq:alpha2}) due to the presence of the $m_\alpha$-th power on the cumulative density function of $Z$. In that case, the approximation errors incurred by the Edgeworth series expansion tend to be amplified, thus potentially violating the upper bound inequality in (\ref{upPfa1}). We will address this issue by adopting an alternative closed-form approximation using results from extreme value theory (EVT), as described next in Section \ref{sec:EVT}.

\section{Edgeworth series approximation for $\beta_m(h)$, the upper bound on $\Pmd$}
\label{sec:Edge}
For a sufficiently large $m$, the distribution of $Z$ in (\ref{defZ}) can be assumed to be Gaussian in virtue of the CLT. This certainly relaxes the complexity of the problem at hand, and provides an acceptable match with the target distribution. However, the CLT approximation is often too loose for small $m$ or when focusing on the tails of the resulting distribution, as it occurs when dealing with error probabilities (e.g. $\Pmd$ and $\Pfa$). A tighter approximation can be obtained through the following theorem \cite[p.~223]{Cramer}:
\begin{theorem}[{Gram-Charlier type-A approximation}]
\label{th:Edge}
The error between the target distribution $f_Z$ and the CLT approximation can be modeled by a series expansion as follows:
\begin{equation}
\epsilon(\tilde{z})\doteq f_Z(\tilde{z})-\phi(\tilde{z})=\phi(\tilde{z})\sum_{p=3}^\infty \frac{C_p}{p!}H_p\p{\zt}\label{eq:Edgeworth}
\end{equation}
where $H_p(\tilde{z})$ is the Hermite polynomial of degree $p$ and $C_p$ the projection of the target distribution onto $H_p(\tilde{z})$,
\eq{\label{eq:coef}
C_p \doteq \int_{-\infty}^\infty H_p\p{\zt}f_Z(\tilde{z}) d\tilde{z}.
}
\end{theorem}

\begin{col}[Gram-Charlier type-A expansion for $f_Z$]
\label{col:f_Edge}
Using the result in (\ref{eq:Edgeworth}), the pdf of $Z$ can be represented through the following series expansion,
\eq{
f_Z(\tilde{z}) = \phi\p{\zt}\corch{1 + \sum_{p=3}^\infty \frac{C_p}{p!}H_p\p{\zt}}.\label{cor1}
}
\end{col}
\begin{col}[Gram-Charlier type-A expansion for $F_Z$]
\label{col:F_Edge}
Integrating the result in (\ref{eq:Edgeworth}), the cdf of $Z$ can be represented through the following series expansion,
\eq{
F_Z(\tilde{z})= \Phi\p{\zt} - \sigma_Z\phi(\tilde{z})\sum_{p=3}^\infty C_pH_{p-1}\p{\zt}.\label{cor2}
}
\begin{IEEEproof}
See Appendix \ref{app:Edge}.
\end{IEEEproof}
\end{col}

While the results in (\ref{cor1})-(\ref{cor2}) provide a closed-form approximation for $f_Z(z)$ and $F_Z(z)$, it is well-known that the Gram-Charlier approximation may suffer from some instabilities and convergence issues \cite{expansions,Kendall}. In particular, the terms of the infinite series in \eqref{eq:Edgeworth} do not monotonically decrease with increasing the order $p$, thus making the truncation of the asymptotic series a not trivial task. Notwithstanding, these issues can be circumvented by rearranging the error terms so as to provide a series expansion with guaranteed convergence \cite{Cramer}. This rearrangement of terms leads to the so-called Edgeworth series expansion, which consists on grouping the error terms with similar order. This is the case, for instance, of terms $p = 3$, $p = \{4,6\}$ and $p = \{5,7,9\}$. Using this observation, we are now in position to provide a closed-form approximation for the upper bound on $\Pmd$ in (\ref{eq:beta2}).

\begin{prop}[Edgeworth approximation for $\Pmd$]
\label{col:Fz_Edge}
Using the result in (\ref{cor2}), the upper bound on $\Pmd$ in (\ref{eq:beta2}) can be approximated as follows
\begin{eqnarray}
\beta_m(h)&\approx&\beta_{\mathrm{EDG},m}(h) \label{pmdAp1}\\
&=& \Phi(h) - \sigma_Z\phi(h) \sum_{p \in \mathcal{A}} C_{p,1}H_{p-1}(h)\nonumber
\end{eqnarray}
where $\mathcal{A} = \{3,4,6\}$ and $C_{p,1}$ are the Hermite coefficients computed using $f_{Z,1}(z)$ under $\Hu$, and given by 
\eq{\label{eq:coefp}
\begin{aligned}
C_{3,1} &= \frac{\xi_{Z,1}^3 -3\xi_{Z,1}\xi_{Z,1}^2 + 2(\xi_{Z,1})^3}{\sigma_{Z,1}^3},\\
C_{4,1} &= \frac{\xi_{Z,1}^4 - 4\xi_{Z,1}\xi_{Z,1}^3 + 6(\xi_{Z,1})^2\xi_{Z,1}^2 -3(\xi_{Z,1})^4}{\sigma_{Z,1}^4} - 3,\\
C_{6,1} &= 10C_{3,1}^2,
\end{aligned}
}
where $\xi_{Z,1}^k$ is the $k$-th order moment of $Z$, which can be evaluated using the results in Section \ref{secMomentsZ} under $\mathcal{H}_1$. Finally, $\sigma_Z$ is the standard deviation of $Z$ that can be obtained as $\sigma_Z=\sqrt{\xi_Z^2 -(\xi_Z)^2}$.
\end{prop}

\section{Approximation for $\alpha_{\ma}(h)$, \:the upper bound on $\Pfa$}
\label{sec:EVT}

\subsection{Edgeworth series approximation}
A closed-form approximation for the upper bound of $\Pfa$ in (\ref{eq:alpha2}) can similarly be obtained substituting $F_{Z,0}$ by its Edgeworth series expansion, as already done in (\ref{pmdAp1}) for $\Pmd$.

\begin{prop}[Edgeworth approximation for $\Pfa$]
\label{col:Pfa_Edge}
Using the result in (\ref{cor2}), the upper bound on $\Pfa$ in (\ref{eq:alpha2}) can be approximated as follows
\begin{eqnarray}
\alpha_{\ma}\p{h} &\approx& \alpha_{\mathrm{EDG},\ma}\p{h}\label{eq:alphaedg}\\
&=&1-\left[\Phi(h) - \sigma_Z\phi(h) \sum_{p \in \mathcal{A}} C_{p,0}H_{p-1}(h)\right]^{m_\alpha}\nonumber
\end{eqnarray}
where $\mathcal{A} = \{3,4,6\}$ and $C_{p,0}$ are the Hermite coefficients computed using $f_{Z,0}(z)$ under $\Ho$, and given by (\ref{eq:coefp}) replacing $\xi_{Z,1}$ by $\xi_{Z,0}$.
\end{prop}

\subsection{Extreme value theory (EVT) approximation}
Although the Edgeworth series provides a better approximation for the tails of the distribution of $Z$ than the CLT, there is still some mismatch between the tail of the approximation and the true distribution. This mismatch is negligible for the case of approximating $F_{Z,1}$ in (\ref{eq:beta2}), but it is not when approximating the $m_\alpha$-th power of $F_{Z,0}$ in (\ref{eq:alpha2}). The approximation inaccuracies become amplified and the upper bound inequality in (\ref{upPfa1}) is not guaranteed to be preserved anymore. 

In order to circumvent this issue we will formulate an alternative approximation for $\Pfa$ making use of results from extreme value theory (EVT). EVT has historically been linked to the statistical analysis of of floods (i.e. flood frequency analysis), where predicting extreme events is of paramount importance. However, EVT is also widely adopted today in applications dealing with finance, insurance or engineering \cite{EVT}. In the problem at hand, we can understand the upper bound in (\ref{eq:alpha2}) as the probability that none of the $m_\alpha$ trials of $Z$ under $\mathcal{H}_0$ exceeds the threshold $h$. If none of them exceeds the threshold, this is equivalent to say that the maximum of these $m_\alpha$ trials does not exceed it either. Following this rationale we make use of the following theorem.
\begin{theorem}[{Extreme Value Theory}]
\label{th:EVT}
Let $U$ be the maximum of $N$ iid samples of $Z$ whose pdf $f_Z$ exhibits exponentially decreasing tails. Then the cdf of $U$ becomes
\eq{\label{eq:EVT}
F_U(u) \stackrel{N\rightarrow\infty}{=}  \exp\p{-e^{-\gamma\p{x-\delta}}},
}
with
\begin{eqnarray}
\delta &=& F_Z^{-1}\p{1 - \frac{1}{N}},\\
\gamma &=& Nf_Z(\delta).
\end{eqnarray}
\end{theorem}
\begin{IEEEproof}
See \cite[p.~166]{EVT}.
\end{IEEEproof}

Using the result above we can provide an alternative approximation for the upper bound on $\Pfa$ as follows,
\begin{prop}[EVT approximation for $\Pfa$]
\label{col:EVT}
Using the result in Theorem 1, the upper bound on $\Pfa$ in (\ref{eq:alpha2}) can be approximated as follows
\begin{eqnarray}
\alpha_{\ma}\p{h}& \approx& \alpha_{\mathrm{EVT},m_\alpha}\p{h} \label{eq:alphaEVT}\\
&=& 1 - \exp\p{-e^{-\gamma_{\ma}\p{h-\delta_{\ma}}}},\nonumber
\end{eqnarray}
where
\begin{eqnarray}
\delta_{\ma} &=& F_{Z,0}^{-1}\p{1 - \frac{1}{\ma}},\label{eq:EVTconst1}\\
\gamma_{\ma} &=& \ma f_{Z,0}(\delta_{\ma}).\label{eq:EVTconst2}
\end{eqnarray}
\end{prop}
\begin{IEEEproof}
We can rewrite the term $[F_{Z,0}(h)]^{\ma}$ in \eqref{eq:alpha2} as
\eq{\label{eq:prob}
[F_{Z,0}(h)]^{\ma} = \Pi\p{\bigcap_{i=1}^{\ma} Z_i < h} = \Pi\p{ M_{\ma}<h},
}
with $M_{\ma} \doteq \max_{1\leq i\leq \ma} \{Z_i\}$. That is, we can obtain $[F_{Z,0}(h)]^{\ma}$ as the probability that $\ma$ iid samples of $Z$ are below the value $h$, which is equivalent to the probability that the maximum of all $\ma$ samples is below $h$. Thereby, we can make use of EVT for obtaining $[\Fo(h)]^{\ma}$. Since $Z$ is the sum of dependent non-central chi-squared and Gaussian random variables, the distribution of $Z$ has an exponentially decreasing tail and Theorem \ref{th:EVT} is applicable to \eqref{eq:prob}. On the other hand, since the pdf and cdf of $Z$ are unknown, we apply the corresponding Edgeworth expansion in order to use Theorem \ref{th:EVT}, and the proof of Corollary \ref{col:EVT} thus follows.
\end{IEEEproof}

Before concluding this section it is worth noting that for obtaining $\delta_{\ma}$ in \eqref{eq:EVTconst1} we need to evaluate the inverse cdf $F_{Z,0}^{-1}$. However, there is not closed-form for this inverse, and then we have to solve the equation $F_{Z,0}(\delta) = 1 - (1/\ma)$ numerically.

\section{Application to signal-level integrity monitoring in GNSS receivers\label{sec:GNSS}}
This section is intended to assess the goodness of the proposed approximations for the upper bound on $\Pmd$ in (\ref{pmdAp1}) and the upper bound on $\Pfa$ in (\ref{eq:alphaedg}) and (\ref{eq:alphaEVT}) For simplicity, we will refer to the former by $\beta_{\mathrm{EDG}}$ and to the latter by $\alpha_{\mathrm{EDG}}$ and $\alpha_{\mathrm{EVT}}$, respectively.

In order to illustrate the goodness of these approximations, we will focus on the specific application of signal-level integrity monitoring in GNSS receivers. {\em Integrity} refers to the ability of the receiver to guarantee the quality and trust of the received signal, in such a way that critical applications can be safely operated. This involves that signal processing techniques must be implemented to analyze some key performance indicators and to detect abnormal values. While this problem has been widely addressed within the civil aviation community \cite{Loh94}, it has always remained at the realm of position, velocity and time (PVT) observables, where measurements from several sources need to be compared and cross-checked for consistency purposes. In the recent years there has been an increasing interest in signal-level metrics as early indicators on the presence of integrity threats, since they are directly linked to the physical received signal and they are readily available before PVT observables are computed. This is the case of the signal-to-noise ratio (SNR), the symmetry of the correlation peak at the matched filter output, or the ratio between the maximum and minimum eigenvalues of the spatial correlation matrix in multi-antenna systems.

The problem to be addressed is clearly a TCD one, where the distribution of the signal-level measurements may suddenly change from its nominal value during the time some threat is present (e.g. a jamming signal, multipath reflections, etc.). Among the wide range of possible signal-level integrity metrics, we focus here on the so-called slope asymmetry metric (SAM). This metric is intended to detect the presence of multipath propagation, which is often the cause of bias errors in the navigation solution and becomes one of the major threats for the safe operation of GNSS receivers in urban environments. The SAM metric is obtained by comparing the left and right slopes of the correlation peak at the output of the matched filter \cite{Parro09}. In nominal conditions, the correlation peak should exhibit a nearly symmetric shape, thus leading the SAM metric to be zero mean. However, the right slope of the correlation peak tends to flatten in the presence of multipath, thus causing the SAM metric to exhibit a nonzero mean.

The distribution of the SAM metric was analyzed in \cite{ICL}, where it was shown to follow the Gaussian mean-and-variance change signal model in (\ref{eq:model}). The following range of values were found to be applicable for a representative urban scenario affected by multipath propagation, according to the measurement campaign conducted in \cite{ION}: $\mu_0=10^{-1}$, $\mu_1=2\cdot10^{-1}$, $\sigma_0^2\in[10^{-4},10^{-2}]$ and $\sigma_1^2\in[10^{-3},10^{-2}]$, where SAM samples are provided at sampling time $T_s=1$ second. Without loss of generality, we will use values within these ranges when assessing the goodness of the Edgeworth approximation for the upper bound on $\Pmd$ in Section \ref{goodPmd}. Next, we will follow with the upper bound on $\Pfa$ in Section \ref{goodPfa}, where both the Edgeworth and EVT approximations will be compared.

Apart from the computation of the corresponding probabilities, we will also evaluate the distance between the exact bound and each of the proposed approximations. This will be done using the Cram\'{e}r-von Mises distance \cite{anderson62},
\begin{equation}
D_{CVM}^2(F,\widehat{F})\doteq\int_{-\infty}^{\infty}\left(F(h)-\widehat{F}(h)\right)^2dF(h)
\end{equation}
where $F(h)$ stands for the exact bound distribution, $\alpha_{\ma}(h)$ or $\beta_m(h)$, and $\widehat{F}(h)$ stands for the proposed approximations. The normalized distance is here computed following \cite{williams08}.

\begin{figure}[t]
\centerline{\includegraphics[width=8.4cm]{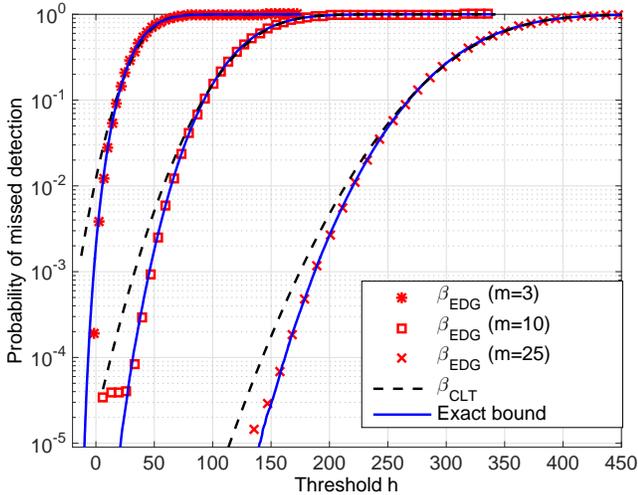}}
\caption{Comparison between the exact bound for the probability of missed detection $\beta_m(h)$ with the corresponding Edgeworth and the CLT approximation for $m = \{3,10,25\}$.}
\label{fig:beta}
\end{figure}

\subsection{Goodness of the upper bound on $\Pmd$\label{goodPmd}}
Fig. \ref{fig:beta} shows the comparison between the exact bound $\beta_m(h)$ in (\ref{eq:beta2}) and both CLT and Edgeworth approximations for different values of the transient duration, $m=\{3,10,25\}$ samples. The signal parameters for the mean-and-variance change signal model are $\muso = 10^{-1}$, $\sigsso = 4\cdot10^{-4}$, $\musu=2\cdot10^{-1}$, $\sigssu =1.6\cdot10^{-3}$. The results in Fig. \ref{fig:beta} show a tight match between the proposed Edgeworth approximation and the exact bound, even for low values of $m$, in contrast to what happens with the CLT approximation. The tight match is particularly true for moderate values of probability of miss down to $10^{-4}$, which comprise the region where integrity techniques typically operate. Some inaccuracies are observed for the Edgeworth approximation, but they are restricted to low values of $m$ and $\beta_m(h) < 10^{-4}$.

\begin{figure}[t]
  \centerline{\includegraphics[width=8.2cm]{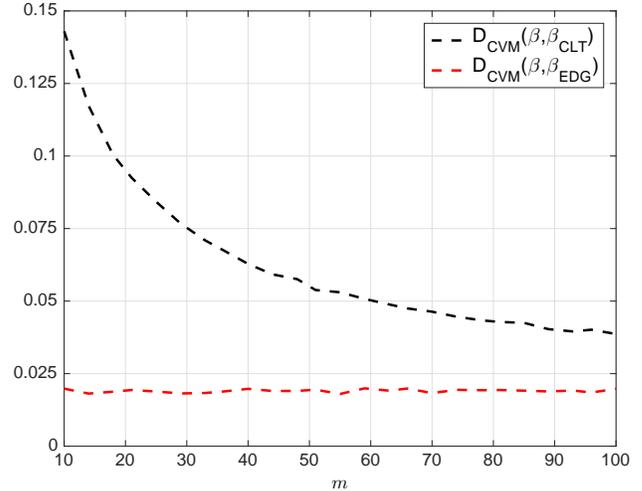}}
\caption{Cram\'{e}r-von Mises distance between the exact bound for the probability of missed detection $\beta$ and the Egeworth and CLT aproximations $\{\beta_{\mathrm{CLT}},\bedg\}$ for the range of interest $0\leq\beta\leq 0.1$.}
\label{fig:KLDPmd}
\end{figure}

For the same signal parameters, the Cram\'{e}r-von Mises distance between the exact bound and the proposed approximations is depicted in Fig. \ref{fig:KLDPmd} as a function of the transient length. The results were obtained for $0\leq\beta\leq 0.1$, which is the range of miss detections that are typically considered in most integrity applications. As can be observed in Fig. \ref{fig:KLDPmd}, the Edgeworth approximation is consistently providing a better match to the exact distribution, when compared to the CLT. It is true though that the accuracy of the CLT approximation improves with the transient length, as more terms are accumulated in (\ref{defSn}). However, the transient would need to be on the order of a few hundred samples length for the CLT to provide similar results to the Edgeworth approximation in terms of probability of miss detection. 

\subsection{Goodness of the upper bound on $\Pfa$\label{goodPfa}}
In this case we have carried out two experiments: the first one for different values of the transient length $m=\{3,10,25,50\}$ and $\ma=10m$ as the time window for guaranteed $\Pfa$; the second one for a fixed transient length $m=6$ and two possible time windows for guaranteed $\Pfa$, namely $\ma=\left\{60,900\right\}$. The latter correspond to 1 and 15 minutes time windows commonly adopted in some integrity applications.

\begin{figure}[t]
  \centerline{\includegraphics[width=8.2cm]{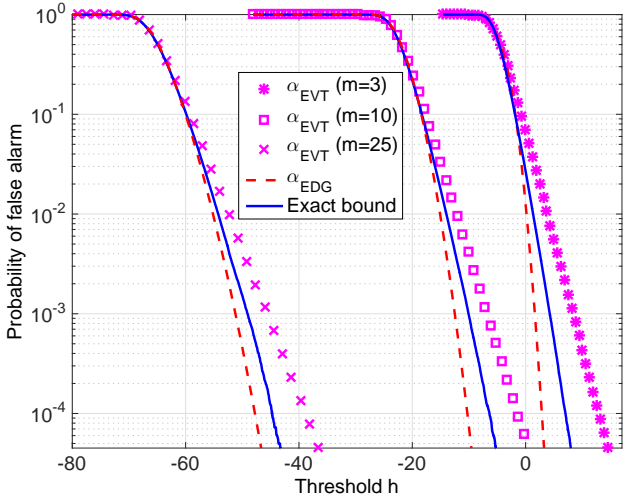}}
\caption{Comparison between the exact bound for the probability of false alarm $\alpha_{\ma}(h)$ with the corresponding Edgeworth and the EVT approximation for $m=\{3,10,25\}$ and $\ma=10\cdot m$.}
\label{fig:alpha}
\end{figure}

The results for the first experiment are shown in Fig. \ref{fig:alpha}, where we see that the match between the Edgeworth approximation and the exact bound $\alpha_{\ma}(h)$ is not that tight as the one previously discussed in Section \ref{goodPmd} for $\beta_m(h)$. Indeed, we now see that there are values of the threshold $h$ where the Edgeworth approximation is actually below the exact bound, thus violating the upper-bound inequality in (\ref{upPfa1}). This is indeed the main reason for the proposed EVT approximation, which is shown in Fig. \ref{fig:alpha} to always remain above the exact bound, thus fulfilling the upper-bound inequality.

If we examine the behavior of both the Edgeworth and EVT approximations as a function of $m$, we can see that the greater the value of $m$, the slightly closer both approximations are to the exact bound. For the Edgeworth approximation, this behavior is due to the fact that the cdf is improved as $m$ increases. However, since we have to compute the $\ma$-th power of this approximation in (\ref{eq:alpha2}), the error terms are still amplified and they cause the overall approximation to violate the inequality in (\ref{upPfa1}). This effect is not observed when using the EVT approximation, which is directly approximating the $\ma$-th power of the cdf and turns out to preserve the upper-bound inequality in (\ref{eq:alpha2}).

\begin{figure}[t]
  \centering
\centerline{\includegraphics[width=8cm]{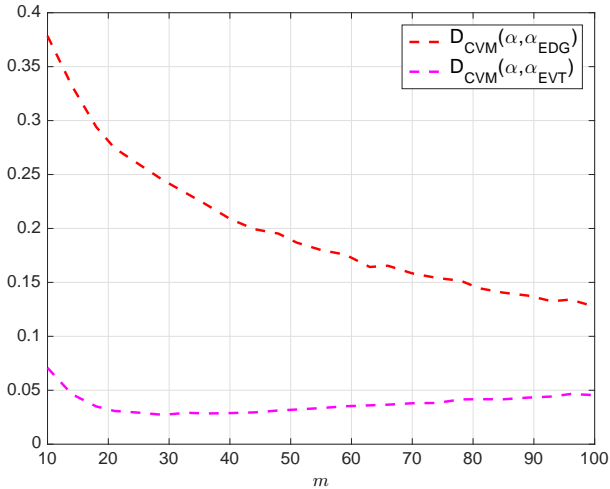}}
\caption{Cram\'{e}r-von Mises distance between the exact bound for the probability of false alarm $\alpha$ and the Edgeworth and EVT aproximations $\{\alpha_{\mathrm{EDG}},\alpha_{\mathrm{EVT}}\}$ for the range of interest $0\leq\alpha\leq 0.1$.}
\label{fig:KLDPfa}
\end{figure}

Fig. \ref{fig:KLDPfa} shows the Cram\'{e}r-von Mises distance between the exact bound $\alpha_{\ma}$ and the Edgeworth and EVT approximations $\{\aedg,\aevt\}$, as a function of $m$. Since we are always interested in having low probabilities of false alarm, the results in Fig. \ref{fig:KLDPfa} have been computed within the range of values $0\leq\alpha\leq 0.1$ (i.e. focusing on the tails of the distributions under analysis). As we can see, the EVT approximation is always providing the closest match to the exact bound, with a quite constant behaviour as a function of $m$. The results for the Edgeworth approximation tend to improve for large values of $m$, due to the larger accumulation of terms in (\ref{defSn}). Nevertheless, the overall distance with respect to the exact bound is still larger than the one achieved by EVT.

\begin{figure}[h]
  \centerline{\includegraphics[width=8.2cm]{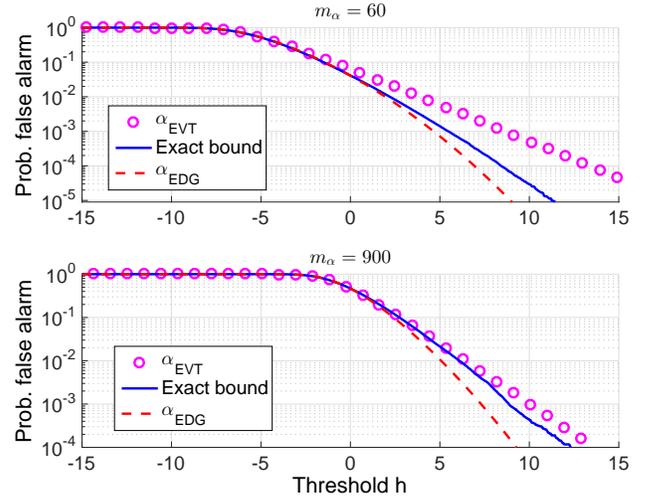}}
\caption{Comparison between the exact bound for the probability of false alarm $\alpha_{\ma}(h)$ with the corresponding Edgeworth and EVT approximation for $\ma = \{60,900\}$ (up and down) and fixed $m=6$.}
\label{fig:alpha_ma}
\end{figure}

The results for the second experiment with fixed $\ma$ corresponding to a 1 and 15 minutes time window are shown in Fig. \ref{fig:alpha_ma}. We see in the upper plot that for $\ma=60$ the Edgeworth approximation is closer to the exact bound than the EVT one, even in the tail. Nonetheless, the Edgeworth approximation is below the exact bound for $h>2$. As we have already mentioned, this is an undesirable behavior that prevents us from using this approximation for upper-bounding $\Pfa$. On the other hand, we see in the lower plot how for a larger value, $\ma=600$, the Edgeworth approximation provides worse results due to the impact of the $\ma$-th power, whereas the EVT approximation actually improves when increasing $\ma$ for a fixed $m$.

\subsection{Performance assessment of the FMA stopping test\label{goodROC}}

\begin{figure}[tb]
  \centering
  \centerline{\includegraphics[width=8cm]{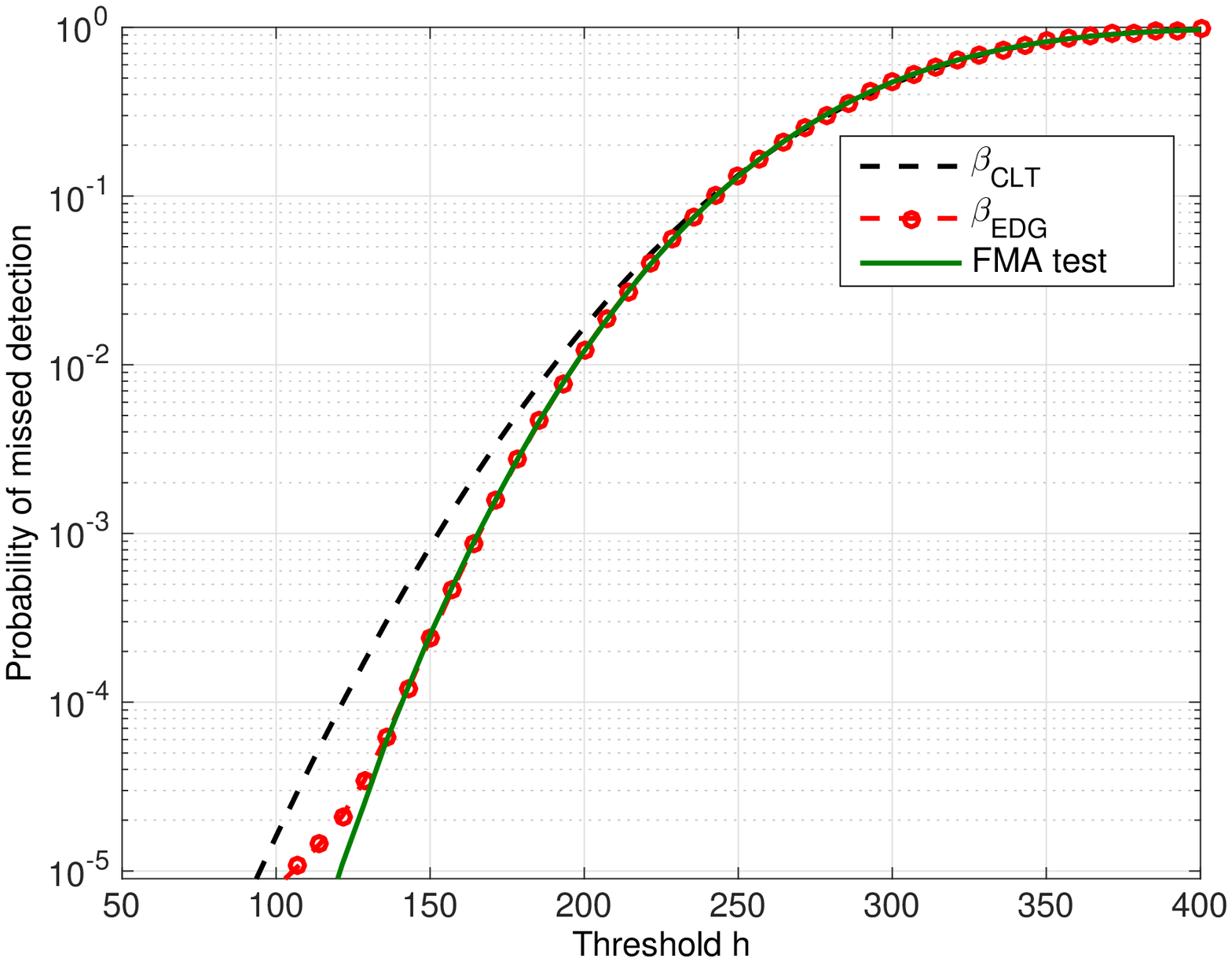}}
\caption{Comparison between the true $\Pmd(\Tf)$ and the proposed approximations for the FMA stopping time test.}
\label{fig:performancePmd}
\end{figure}

Once the goodness of the proposed approximations for $\beta_m(h)$ and $\alpha_{\ma}(h)$ has been presented, we can now proceed with the performance assessment of the FMA test. To do so, we have numerically evaluated the true $\Pmd(\Tf)$ and $\Pfa(\Tf)$ for the FMA test in (\ref{eq:Tfma}), and we have compared the resulting values with the proposed approximations. The results can be observed first in Fig. \ref{fig:performancePmd} for $\Pmd$ as a function of the detection threshold $h$. We can see how the CLT approximation clearly departs from the true $\Pmd(\Tf)$ for probabilities smaller than $10^{-2}$. In contrast, the Edgeworth approximation provides a really tight match with the true $\Pmd(\Tf)$ for values down to $10^{-4}$, thus providing a two-orders-of-magnitude improvement with respect to the CLT.

\begin{figure}
  \centering
  \centerline{\includegraphics[width=8cm]{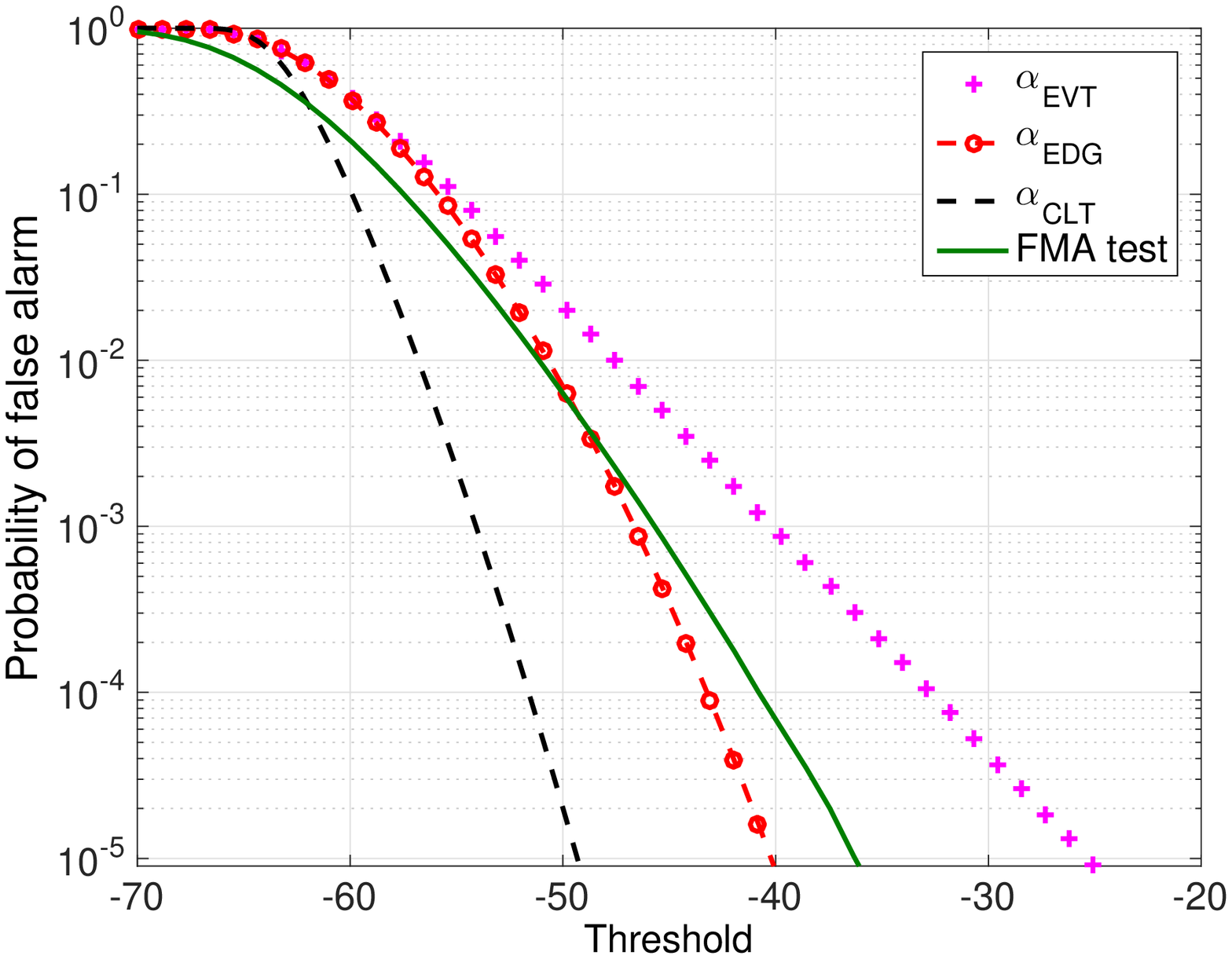}}
\caption{Comparison between the true $\Pfa(\Tf)$ and the proposed approximations for the FMA stopping time test.}
\label{fig:performancePfa}
\end{figure}

Results for $\Pfa$ are shown in Fig. \ref{fig:performancePfa}, where we see how the Edgeworth approximation fulfills the upper-bound down to a probability of $10^{-2}$, only. In contrast, the EVT approximation permanently upper-bounds $\Pfa(\Tf)$ in the whole range of probabilities. The results have been obtained using the same parameters as for Fig. \ref{fig:beta} but with $m=6$, $\ma = 60$, and $\musu = 0.3$.

\begin{figure}[h]
  \centering
  \centerline{\includegraphics[width=8.2cm]{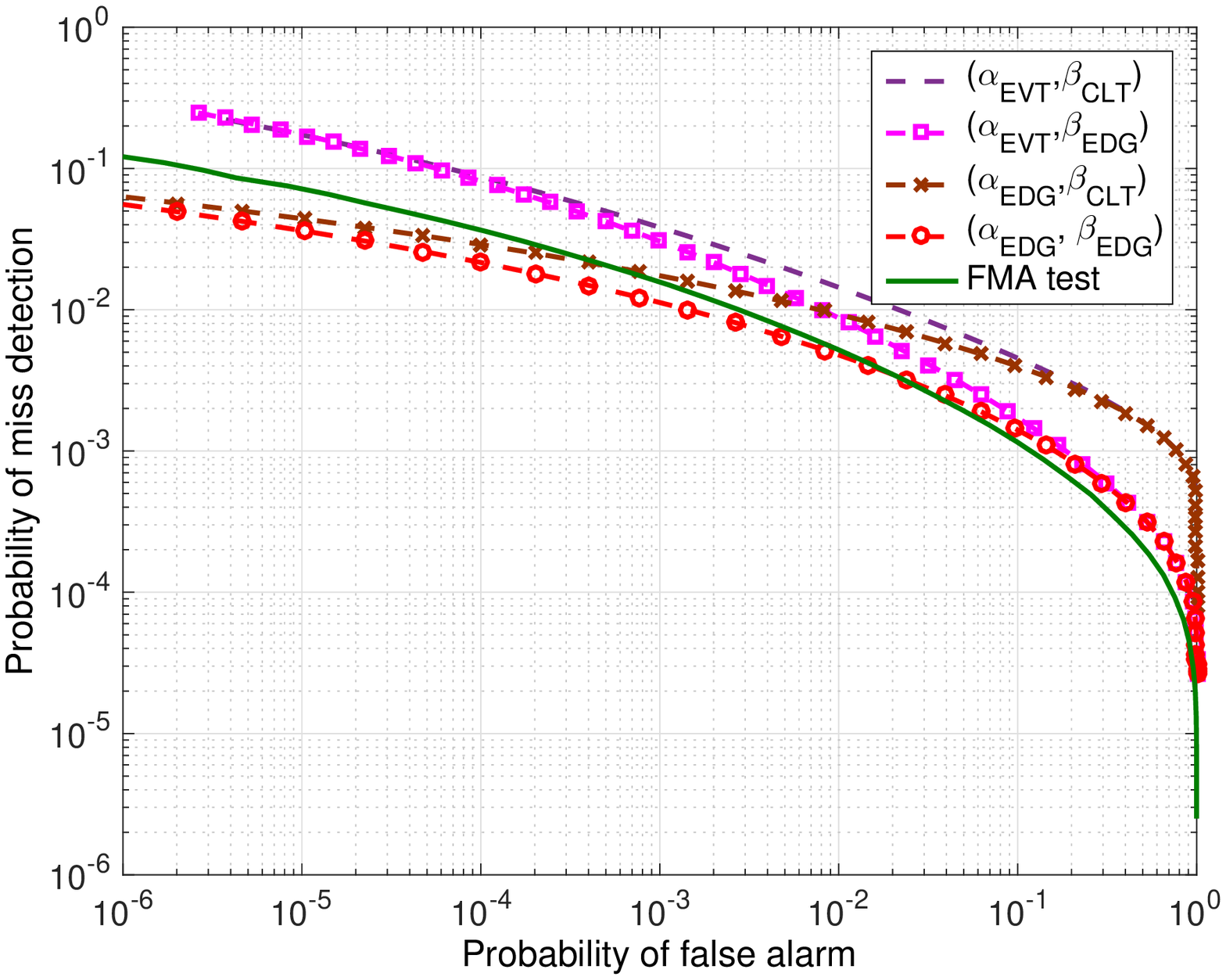}}
\caption{Numerical simulation of the ROC for the SAM-based FMA stopping time. Comparison between the real ROC, the approximated ROC by using either $\aevt$ in \eqref{eq:alphaEVT} or $\aedg$ \eqref{eq:alphaedg} with $\bclt$ and $\bedg$ \eqref{eq:bedg}.}
\label{fig:ROC}
\end{figure}

Finally, we have combined the results from $\Pmd$ and $\Pfa$ to build the receiver operating characteristics (ROC) shown in Fig. \ref{fig:ROC}. Please note that this definition of ROC is different from the standard one, since probability of missed detection is considered here instead of probability of detection. The same parameters as in Fig. \ref{fig:alpha_ma} have been used with $m=6$ and $\ma=300$, and the true (i.e. numerical) results are compared with the different approximations presented so far in this work. That is, the Edgeworth and EVT approximations for the bound on $\Pfa(\Tf)$ (i.e. $\aevt$ and $\aedg$), and the CLT and Edgeworth approximations for the bound on $\Pmd(\Tf)$ (i.e. $\bedg$ and $\bclt$). Results are shown in Fig. \ref{fig:ROC} for all possible pairs of approximations, namely $\left\{\aevt,\bclt\right\}$, $\left\{\aevt,\bedg\right\}$, $\left\{\aedg,\bclt\right\}$ and $\left\{\aedg,\bedg\right\}$.

We are looking here for the best upper bound approximation to the FMA test performance, and for the case under study, the best upper bound is provided by the pair $\left\{\aedg,\bclt\right\}$ as seen in Fig. \ref{fig:ROC}. This is the pair providing the uniformly closest upper bound to the true FMA performance. Some other pairs provide a closer upper bound, but just for a finite range of missed detection or false alarm probabilities. This is therefore a clear example showing the interest in the use of the proposed Edgeworth and EVT approximations for inhomogeneous detection problems, which span outside the domain of the specific application on GNSS signal-level integrity considered herein.

\section{Conclusions}
\label{sec:conclusions}
This work has provided closed-form approximations for the pdf and cdf of inhomogeneous quadratic tests, which are composed of sum of dependent chi-square and Gaussian random variables. These tests typically appear in detection problems where Gaussian samples are subject to a simultaneous change in both mean and variance. We have introduced a transient change detection problem in the context of GNSS signal-level integrity monitoring in order to further illustrate this problem, and to motivate the need for accurate and closed-form expressions to assess the detection performance. To this end, we have proposed two simple and closed-form approximations for the upper bounds on the probability of missed detection and probability of false alarm, which solve the problem of the unknown statistical distribution of inhomogeneous quadratic forms. Simulation results have been obtained using realistic parameters in order to assess the goodness of the proposed approximations, and the superior performance with respect to the widely adopted approximation relying on the CLT. While the application was kept in the context of GNSS, the results are general and could be applied to any other field where inhomogeneous quadratic tests need to be implemented.

\appendices

\section{Proof of Corollary \ref{col:F_Edge}}
\label{app:Edge}
\begin{lemma}
Let $H_p(\tilde{z})$ be the Hermite polynomial of degree $p$, then
\eq{\label{eq:derivative}
\p{\frac{d}{d\tilde{z}}}^p\phi(\tz) = (-1)^p H_p(\tz)\phi(\tz).
}
\end{lemma}
\begin{IEEEproof}
First, note that $\phi(\tz) \doteq e^{-\tz^2/2}/\sqrt{2\pi}$, and then
\eq{\label{eq:exps}
\begin{aligned}
e^{-\tz^2/2} &= \sqrt{2\pi}\phi(\tz), \\
e^{\tz^2/2} &= \p{\sqrt{2\pi}\phi(\tz)}^{-1}.
\end{aligned}
}
On the other hand, the Hermite polynomials are defined as
\eq{\label{eq:Hermite}
H_p(\tz) \doteq (-1)^p e^{\tz^2/2}\p{\frac{d}{d\tz}}^p e^{-\tz^2/2}.
}
Hence, substituting \eqref{eq:exps} into \eqref{eq:Hermite} we obtain
\eq{
H_p(\tz) = (-1)^p(\phi(\tz))^{-1}\p{\frac{d}{d\tz}}^p \phi(\tz),
}
which leads to \eqref{eq:derivative}.
\end{IEEEproof}

The proof of Corollary \ref{col:f_Edge} follows straight away from the Taylor series expansion of $f_Z(\tz)$ and the orthogonal properties of the Hermite polynomials \cite{Cramer}. However, some further manipulations are required to proof Corollary \ref{col:F_Edge}. Starting from the cdf definition, we have that
\eq{\label{eq:cdfint}
\begin{aligned}
F_Z(\tz) &\doteq \int_{-\infty}^{\tz} f_Z(u) du \\
&\approx \Phi(\tilde{z}) + \sum_{p=3}^\infty \frac{C_p}{p!} \int_{-\infty}^{\tilde{z}} \phi(\tilde{u})H_p(\tilde{u}) du,
\end{aligned}
}
where $\tilde{u}\doteq(u-\mu_Z)/\sigma_Z$ and the first term follows by the definition of the standard Gaussian cdf. The integral is solved by integrating \eqref{eq:derivative},
\eq{
\begin{aligned}
\int_{-\infty}^{\tz} \phi(\tilde{u})H_p(\tilde{u}) du &= \sigz \int_{-\infty}^{\tilde{z}} (-1)^p\p{\frac{d}{d\tilde{u}}}^p\phi(\tilde{u}) d\tilde{u} \\
&= \sigz (-1)^p \corch{\p{\frac{d}{d\tilde{u}}}^{p-1}\phi(\tilde{u})}_{-\infty}^{\tilde{z}},
\end{aligned}
}
where the first equality follows by applying a change of variable. Applying \eqref{eq:derivative} we have
\eq{
\begin{aligned}
\int_{-\infty}^{\tz} \phi(\tilde{u})H_p(\tilde{u}) du 	&= \sigma_Z (-1)^{2p-1}H_{p-1}(\tilde{z})\phi(\tilde{z})\\
									&= -\sigz H_{p-1}(\tilde{z})\phi(\tilde{z}),
\end{aligned}
}
and then (\ref{cor2}) follows by substituting this result into \eqref{eq:cdfint}.

\bibliographystyle{IEEEtran}
\bibliography{biblio}
\end{document}